\documentclass[12pt,reqno]{amsart}

\usepackage{amsthm,amssymb,amstext,amscd,euscript,mathrsfs,amsfonts,amsbsy,amsxtra,latexsym,amsmath}
\usepackage{fullpage}
\usepackage[english]{babel}
\usepackage[latin1]{inputenc}
\usepackage{verbatim}
\usepackage{graphicx} 
\usepackage[all]{xy} 
\usepackage{enumitem}
\usepackage{bbm}
\usepackage{marvosym}
\numberwithin{figure}{section} 
\allowdisplaybreaks

\newcommand{\field}[1]{\mathbb{#1}}

\newcommand{\C}{\field{C}}

\numberwithin{equation}{section}

\numberwithin{theorem}{section}

\theoremstyle{definition}

\newtheorem{remark}{Remark}[section]

\newcommand{\bea}{\begin{eqnarray}} 
\newcommand{\eea}{\end{eqnarray}} 
\newcommand{\be}{\begin{equation}} 
\newcommand{\ee}{\end{equation}} 
\newcommand{\benn}{\begin{equation*}} 
\newcommand{\eenn}{\end{equation*}}

\title[short title]{Orientifold limits of singular $F$-theory vacua}
\author{James Fullwood}
\address{School of Mathematical Sciences\\ Shanghai Jiao Tong University \\ 800 Dongchuan Road, Shanghai, China}
\email{fullwood@sjtu.edu.cn}
\author{Dongxu Wang}
\address{Department of Mathematics\\Dongbei University of Finance and Economics\\ 217 Jianshan St, Shahekou District, Dalian, China}
\email{dxwang@dufe.edu.cn}
\begin{document}

\maketitle

\begin{abstract}
We construct global orientifold limits of singular $F$-theory vacua whose associated gauge groups are SO(3), SO(5), SO(6), $F_4$, SU(4), and Spin(7). For each limit we show a universal tadpole relation is satisfied, which is a homological identity whose dimension-zero component encodes the matching of the D3 charge between each $F$-theory compactification and its orientifold limit. While for smooth $F$-theory compactifications which admit global orientifold limits the contribution to the associated universal tadpole relation comes from its Chern class, we show that for all singular $F$-theory compactifications under consideration, the contribution to the universal tadpole relation comes from its \emph{stringy} Chern class. Moreover, the brane spectrum associated with each of our limits consists solely of smooth branes, as opposed to the Sen limit which necessarily involves singular branes.
\end{abstract}

\section{Introduction}
$F$-theory compactifications of string vacua are related to weakly coupled type-IIB compactifications via S-duality, which relates strongly coupled regimes with weakly coupled ones. For $F$-theory compactified on an elliptic Calabi-Yau $(n+1)$-fold $X\to B$ whose total space may be given by a global Weierstrass equation, Sen was able to identify a certain limit in the complex structure moduli space of such compactifications with an orientifold theory compactified on an $n$-fold which is the total space of a ramified double cover $Z\to B$ \cite{Sen}. In particular, the $j$-invariant of the elliptic fibers in the limit constructed by Sen generically approach infinity, signaling weak coupling almost everywhere on $B$, and a monodromy analysis of the associated limiting discriminant reveals the presence of an O7-plane and a D7-brane (for $n=3$).  

A non-trivial consistency condition for an orientifold limit of $F$-theory consists of a comparison of the D3 tadpole between the two theories in the absence of fluxes, which should be equal. For $F$-theory compactified on a general elliptic Calabi-Yau 4-fold $X\to B$, we have that the D3 tadpole is given by
\[
N_{D3}=\frac{1}{24}\chi(X),
\] 
where $\chi(X)$ denotes the topological Euler characteristic of $X$, while on the type-IIB side we have
\[
N_{D3}=\frac{1}{2}\left(4\sum_i \frac{\chi(O_i)}{24}+\sum_j\frac{\chi(D_j)}{24}\right),
\]
where $O_i$ and $D_j$ denote the supports of the O7-planes and D7-branes. As such, in an orientifold limit of $F$-theory, we expect to find
\begin{equation}\label{TR}
2\chi(X)=4\sum_i \chi(O_i)+\sum_j \chi(D_j).
\end{equation}

In the limit constructed by Sen, the brane spectrum consists of an O7-plane and a single D7-brane, which wraps a singular divisor $D$ in the total space of the orientifold $Z\to B$, whose equation takes the form 
\[
D:(\eta^2+12\zeta^2\vartheta=0) \subset Z.
\]
The surface $D$ then acquires singularities along the curve $\eta=\zeta=0$, which enhance to pinch-point singularities along the zero-dimensional locus $\eta=\zeta=\vartheta=0$. One must then be careful how to incorporate the charge of $D$ into the D3 tadpole, since in string theory, classical invariants of singular varieties must often be replaced with their `stringy' versions. In the case of Sen's limit, it was first shown in \cite{AE1} that the modification of the D3 tadpole constraint \eqref{TR} due to the singularities of $D$ takes the form
\begin{equation}\label{E8TP}
2\chi(X)=4\chi(O)+\chi_{\text{str}}(D)-\chi(S),
\end{equation}
where $O$ denotes the O7-plane, $\chi_{\text{str}}(D)$ denotes the \emph{stringy} Euler characteristic of $D$, and $S$ denotes the pinch-locus of $D$. A physical argument for $\chi_{\text{str}}(D)-\chi(S)$ being the contribution of $D$ to the D3 tadpole on the type-IIB side was provided in \cite{CDM}, while a top-down derivation of the tadpole constraint \eqref{E8TP} using only mathematical considerations was provided in \cite{FTR2}. Further insights into the dictionary between $F$-theory and Sen's limit were also provided in \cite{SL}, and other foundational work on orientifold compactifications in $F$-theory may be found in \cite{Buchel}\cite{Morrison-Vafa1}\cite{Morrison-Vafa2}\cite{Vafa}.

From a purely mathematical perspective, a compelling aspect of the D3 tadpole constraint \eqref{E8TP} associated with Sen's limit, is that it is the dimension-zero component of a homological identity which holds in a much broader context than considered by physicists \cite{AE1}. In particular, if we denote the projection of the $F$-theory elliptic fibration by $\varphi:X\to B$, and the orientifold projection by $\rho:Z\to B$, then equation \eqref{E8TP} is the dimension-zero component of the homological Chern class identity given by
\begin{equation}\label{E8UTR}
2\varphi_*c(X)=\rho_*(4c(O)+c_{\text{str}}(D)-c(S)),
\end{equation}
where $c_{\text{str}}(D)$ denotes the stringy Chern class of $D$. Moreover, not only does the identity \eqref{E8UTR} hold over a base $B$ of arbitrary dimension, it also holds without any Calabi-Yau hypothesis on $X$ (as the only requirement on $X$ is that it may be given by a global Weierstrass equation). As such, equation \eqref{E8UTR} is often referred to as the \emph{universal tadpole relation} associated with Sen's limit. Universal tadpole relations for orientifold limits of smooth $F$-theory vacua which do not admit global Weierstrass equations were also shown to hold in \cite{AE2}\cite{EFY}\cite{CCG}, and a universal tadpole relation associated with an oriented type-IIB limit was shown to hold in \cite{SLFW}.

The fact that Sen's limit is defined exclusively for smooth Weierstrass fibrations $X\to B$ is quite restrictive. On the $F$-theory side, this prohibits one from geometrically engineering non-abelian gauge theories, as one may associate a non-abelian gauge theory with a Weierstrass fibration only once singularities are introduced into the total space of the fibration. As such, the state of the art in engineering non-abelian gauge theories associated with an $F$-theory compactification,  is to work with the \emph{Tate form} of singular Weierstrass fibrations $X\to B$ \cite{TFoS}, which is given by
\begin{equation}\label{tf}
y^2z + a_1xyz + a_3yz^2 = x^3 + a_2x^2z + a_4xz^2 + a_6z^3.
\end{equation}    
The Tate form of a Weierstrass fibration is particularly useful due to the fact that given a Kodaira fiber $\mathfrak{f}$, Tate's algorithm provides a precise recipe for tuning the coefficients $a_i$ in such a way that $\mathfrak{f}$ will appear over a divisor $S\subset B$ upon a resolution of singularities $\widetilde{X}\to X$ \cite{Tate}. The fiber $\mathfrak{f}$ together with the Mordell-Weil group of $X$ then determines the associated gauge group. In light of this, Sen's limit was generalized by Donagi and Wijnholt in \cite{DWHBUV} to singular Weierstrass fibrations in Tate form. Weak coupling limits of other singular compactifications were also investigated in \cite{KMW}\cite{BCV}.

The orientifold limits of singular Weierstrass fibrations constructed by Donagi and Wijnholt are used in their analysis to construct local models, thus global constraints such as the D3 tadpole are never considered in such limits. Furthermore, as pointed out by Esole and Savelli \cite{ESTF}, it is often the case with such limits (e.g. limits of $\text{SU}(n)$ theories) that the associated orientifold admits conifold singularities whose crepant resolutions are not compatible with the orientifold involution. In this note, we then construct global orientifold limits of singular $F$-theory vacua which admit the gauge groups SO(3), SO(5), SO(6), $F_4$, SU(4), and Spin(7). Outside of the SO(3) case each of our limits are distinct from the Donagi-Wijnholt limits, and we show that the D3 tadpole constraint given by \eqref{TR} holds in each of our limits once $\chi(X)$ is replaced by the stringy Euler characteristic $\chi_{\text{str}}(X)$. This yields compelling evidence that the D3 tadpole associated with a singular $F$-theory compactification is given by
\[
N_{D3}=\frac{1}{24}\chi_{\text{str}}(X),
\] 
as opposed to being proportional to the usual topological Euler characteristic. And just as in the case of Sen's limit (and in the limits constructed in \cite{AE2}\cite{EFY}), we show that each of the numerical tadpole relations are the dimension zero component of a much more general homological identity involving the stringy Chern class $c_{\text{str}}(X)$, which holds in a much broader context than its physical origins. Unlike the Sen limit however, in all the limits we construct the branes which arise in the limits are all supported on smooth divisors, so there is no need to modify the $D3$ charge on the type-IIB side as in Sen's limit. Moreover, the total space of the orientifold in all of our limits are smooth, so there is no issue with whether or not a crepant resolution is compatible with the orientifold involution. We also note that the $F_4$ and SO(6) cases admit the same limit, as do the SU(4) and Spin(7) cases, and as such, these cases provide distinct $F$-theory lifts of their orientifold limits. Furthermore, the SO(6) and $F_4$ limits provide the first geometric realizations of a universal tadpole relation that was shown to exist in \cite{AE2} (i.e., the $(1,...,1)$ case of Theorem~4.9). While all of our limits involve smooth branes on the type-IIB side, we show in \S\ref{SSB} that for certain models of more phenomenological interest -- such as SU(5) for example -- a universal tadpole relation involving solely smooth branes is not possible.  

As the Weiertrass fibrations we consider are all singular, either one defines the $F$-theory compactification on a crepant resolution $\widetilde{X}\to X$, or one takes up the issue of defining $F$-theory on $X$ itself \cite{CSSF}. While the language used in this note tends to favor the latter approach, we note that our results may be adapted to the former approach as well, since stringy invariants of $X$ coincide with those of $\widetilde{X}$.

\section{The singular $F$-theory vacua under consideration}\label{VUC}
Let $B$ be a compact complex manifold, $\mathscr{L}\to B$ a holomorphic line bundle, and let $\mathscr{E}\to B$ be the rank 3 vector bundle given by
\begin{equation}\label{VB}
\mathscr{E}=\mathscr{O}_B\oplus \mathscr{L}^2\oplus \mathscr{L}^3,
\end{equation}
where $\mathscr{L}^k$ denotes the $k$th tensor power of $\mathscr{L}$. We then consider the projective bundle of \emph{lines} in $\mathscr{E}\to B$, which is a $\mathbb{P}^2$-bundle $\pi:\mathbb{P}(\mathscr{E})\to B$ whose tautological bundle we denote by $\mathscr{O}(-1)$.  The vacua we consider are all elliptic fibrations $X\to B$, whose total space is a hypersurface in $\mathbb{P}(\mathscr{E})$ which may be given by a global Weierstrass equation
\[
X:(y^2z=x^3+Fxz^2+Gz^3)\subset \mathbb{P}(\mathscr{E}).
\]
In the equation for $X$, $x$, $y$ and $z$ are projective coordinates on the fibers of $\pi:\mathbb{P}(\mathscr{E})\to B$, which are sections of $\mathscr{O}(1)\otimes \pi^*\mathscr{L}^2$, $\mathscr{O}(1)\otimes \pi^*\mathscr{L}^3$ and $\mathscr{O}(1)$ respectfully. The coefficients $F$ and $G$ are then sections of $\pi^*\mathscr{L}^4$ and $\pi^*\mathscr{L}^6$, so that $X$ corresponds to the zero-scheme of a section of $\mathscr{O}(3)\otimes \pi^*\mathscr{L}^6$. The singular fibers of $X\to B$ lie over the discriminant of $X$, which is given by
\[
\Delta: (4F^3+27G^2=0)\subset B,
\] 
so that $\Delta$ is the zero-scheme of a section of $\mathscr{L}^{12}$. Over a general point of the discriminant the fibers are nodal cubics, which then enhance to cuspidal cubics along $F=G=0$. When $\mathscr{L}$ is the anti-canonical bundle $\mathscr{O}(-K_B)$, the canonical class $K_X$ is trivial, thus the case $\mathscr{L}=\mathscr{O}(-K_B)$ with $B$ of dimension 2 or 3 is the case of physical interest. However, no such assumptions are needed for our calculations. 

For our purposes it will be more useful to work with the \emph{Tate form} of $X$, which is obtained from the Weierstrass equation from a linear change of coordinates. In particular, the Tate form of $X$ is given by
\[
X:(y^2z+a_1xyz+a_3yz^2=x^3+a_2x^2z+a_4xz^2+a_6z^3)\subset \mathbb{P}(\mathscr{E}).
\]
In the Tate form of $X$, each $a_i$ is a section of $\pi^*\mathscr{L}^i$, and with regards to the Weiertsrass form of $X$, we have
\[
F=-\frac{1}{48}(b_2^2-24b_4), \quad 
G=-\frac{1}{864}(36b_2b_4 - b_2^3 - 216b_6),
\]
where
\[
b_2 = a_1^2 + 4a_2, \quad
b_4 = a_1a_3 + 2a_4, \quad \text{and} \quad 
b_6 = a_3^2 + 4a_6.
\]
From here on, we will denote $\pi^*\mathscr{L}^i$ simply by $\mathscr{L}^i$ for ease of notation.

Each $X$ we consider admits singularities over a smooth divisor $S\subset B$ which determines an associated gauge group $\mathcal{G}_X$ (to see how the singularities of $X$ determine $\mathcal{G}_X$, one may consult for example \cite{EJK}). In particular, we consider $X$ whose associated gauge groups are SO(3), SO(5), SO(6), $F_4$, SU(4) and Spin(7). After denoting a regular section of $\mathscr{O}(S)$ by $s$, the singular locus of each fibration is given by $x=y=s=0$. 

\subsection{SO(3) fibrations} 
For $X\to B$ an SO(3) fibration, the Tate form is given by
\[
\text{SO}(3): y^2z=x^3+a_2x^2z+sxz^2,
\]
so that $a_4=s$ in this case. Note that this constrains $s$ to be a section of $\mathscr{L}^4$. The discriminant of SO(3) fibrations is then given by
\[
\Delta_{\text{SO}(3)}: s^2(a_{2}^2 - 4s)=0.
\]

\subsection{SO(5) fibrations}
For $X\to B$ an SO(5) fibration, the Tate form is given by
\[
\text{SO}(5): y^2z=x^3+a_2x^2z+s^2xz^2,
\]
so that $a_4=s^2$ in this case. Note that this constrains $s$ to be a section of $\mathscr{L}^2$. The discriminant of SO(3) fibrations is then given by
\[
\Delta_{\text{SO}(5)}: s^4(a_{2} - 2s)(a_{2} + 2s)=0.
\]

\subsection{SO(6) fibrations}
For $X\to B$ an SO(6) fibration, the Tate form is given by
\[
\text{SO}(6): y^2z+a_1xyz=x^3+sx^2z+s^2xz^2,
\]
so that $a_4=s^2$ and $a_2=s$ in this case. Note that this constrains $s$ to be a section of $\mathscr{L}^2$. The discriminant of SO(6) fibrations is then given by
\[
\Delta_{\text{SO}(6)}: s^4(a_{1}^2 - 4s)(a_{1}^2 + 12s)=0.
\]

\subsection{$F_4$ fibrations}
For $X\to B$ an $F_4$ fibration, the Tate form is given by
\[
F_4: y^2z=x^3+c_1s^3xz^2+c_2s^4z^3,
\]
so that $a_4=c_1s^3$ and $a_6=c_2s^4$ in this case. Note that this constrains $s$ to be a section of $\mathscr{L}$, while $c_i$ is constrained to be a section of $\mathscr{L}^i$. The discriminant of $F_4$ fibrations is then given by
\[
\Delta_{F_4}: s^8(4sc_1^3+27c_2^2).
\]

\subsection{SU(4) fibrations}
For $X\to B$ an SU(4) fibration, the Tate form is given by
\[
F_4: y^2z+a_1xyz=x^3+c_1sx^2z+c_2s^2xz^2+d_2s^4z^3,
\]
so that $a_2=c_1s$, $a_4=c_2s^2$ and $a_6=d_2s^4$ in this case. Note that this constrains $s$ to be a section of $\mathscr{L}$, while $c_i$ and $d_i$ are constrained to be a sections of $\mathscr{L}^i$. The discriminant of SU(4) fibrations is then given by
\[
\Delta_{\text{SU(4)}}: s^4(a_1^6d_2+12a_1^4c_1d_2s-a_1^4c_2^2+48a_1^2c_1^2d_2s^2-8a_1^2c_1c_2^2s-72a_1^2c_2d_2s^2+64c_1^3d_2s^3
\]
\[
-16c_1^2c_2^2s^2-288c_1c_2d_2s^3+64c_2^3s^2+432d_2^2s^4).
\]

\subsection{Spin(7) fibrations}
For $X\to B$ a Spin(7) fibration, the Tate form is given by
\[
\text{Spin(7)}: y^2z=x^3+c_1sx^2z+c_2s^2xz^2+d_2s^4z^3,
\]
so that $a_4=c_1s^3$ and $a_6=c_2s^4$ in this case. Note that this constrains $s$ to be a section of $\mathscr{L}$, while $c_i$ and $d_i$ are constrained to be a sections of $\mathscr{L}^i$. The discriminant of Spin(7) fibrations is then given by
\[
\Delta_{\text{Spin(7)}}: s^6(4c_1^3d_2s-c_1^2c_2^2-18c_1d_2c_2s+4c_2^3+27d_2^2s^2).
\]

\section{Orientifold limits}\label{OL}
In this section, we review Sen's limit for smooth Weierstrass fibrations, and then construct orientifold limits for the six families of vacua whose equations were given in \S\ref{VUC}. In all cases $B$ denotes a compact complex manifold, $\mathscr{L}\to B$ a line bundle, and $\mathscr{E}$ denotes the rank 3 vector bundle given by \eqref{VB}.

\subsection{Sen's limit revisited}
Let $\psi: W\to B$ be a smooth Weierstrass fibration, so that $W$ is a hypersurface in $\mathbb{P}(\mathscr{E})$ given by
\[
W: y^2z=x^3+fxz^2+gz^3.
\]
As in the case of the singular Weierstrass fibrations introduced in the previous section, $x$, $y$, and $z$ are sections of $\mathscr{O}(1)\otimes \mathscr{L}^2$, $\mathscr{O}(1)\otimes \mathscr{L}^3$ and $\mathscr{O}(1)$ respectively, while $f$ and $g$ are sections of $\mathscr{L}^4$ and $\mathscr{L}^6$ respectively. To ensure that $W$ is smooth, we assume that the hypersurfaces in $B$ given by $f=0$ and $g=0$ are both smooth and intersect transversally. The singular fibers of $W\to B$ lie over the discriminant locus $\Delta\subset B$, which is given by
\[
\Delta:4f^3+27g^2=0.
\] 
In particular, over a generic point of $\Delta$ the fibers are nodal cubics, which then enhance to cuspidal cubics over the codimension 2 locus given by $f=g=0$.
 
The orientifold limit constructed by Sen is then achieved by taking
\[
f=-3h^2+t\eta, \quad g=-2h^3+t h\eta+t^2\vartheta,
\]
where $h$ and $\eta$ are general sections of $\mathscr{L}^2$ and $\mathscr{L}^4$ respectfully, and $t$ is a complex deformation parameter which varies over a disk $\mathscr{D}\subset \mathbb{C}$ centered about the origin. Such redefinitions then give rise to a family $\mathscr{W}\to \mathscr{D}$, whose total space is given by
\[
\mathscr{W}:(y^2z=x^3+(-3h^2+t\eta)xz^2+(-2h^3+t h\eta+t^2\vartheta)z^3)\subset \mathbb{P}(\mathscr{E})\times \mathscr{D}.
\]
The central fiber of the family is then given by
\[
\mathscr{W}_0:(y^2z=x^3-3h^2xz^2-2h^3z^3)\subset \mathbb{P}(\mathscr{E}),
\]
which is degenerate fibration with only singular fibers. In particular, the fibers are generically nodal cubics, which signals weak coupling as the $j$-invariant of an elliptic curve approaches $\infty$ as the curve approaches a nodal singularity. The fibers then enhance to cuspidal cubics along the hypersurface $O\subset B$ given by $h=0$, which is then identified as the orientifold hyperplane. We then take a double cover $\rho: Z\to B$ ramified along $O$, which is achieved by defining $Z$ to be the hypersurface in the total space of $\mathscr{L}\to B$ which is given by
\[
Z:(\zeta^2=h)\subset \mathscr{L},
\]
where $\zeta$ is a section (of the pullback to $\mathscr{L}$) of $\mathscr{L}^2$. We summarize the geometry of the situation via the following diagram
\[
\xymatrix{
Z \ar[dr]_{\rho} \ar[r]  & \mathscr{L} \ar[d]^f & \ar[l] \ar[d] f^*\mathscr{L}^2   \\
W \ar[r]_{\psi} & B & \ar[l] \mathscr{L}^2.  \\
}
\]

The orientifold involution is then given by $\zeta \mapsto -\zeta$, and a simple adjunction calculation shows that $K_Z$ is trivial if and only if $K_W$ is, i.e., when $\mathscr{L}=\mathscr{O}(-K_B)$. To arrive at the brane spectrum associated with the limit, we then take the flat limit of the associated family of discriminants as $t\to 0$, and then pull it back to $Z$. In particular, expanding the associated family of discriminants viewed as a function of $t$ yields
\[
\Delta(t)=h^2(\eta^2+12h\vartheta)t^2+\cdots,
\] 
thus the flat limit as $t\to 0$ is given by
\[
\Delta_0: (h^2(\eta^2+12h\vartheta)=0)\subset B.
\]
Pulling $\Delta_0$ back to $Z$ then yields
\[
\overline{\Delta}_0:(\zeta^4(\eta^2+12\zeta^2 \vartheta)=0)\subset Z.
\]
We then see that the brane spectrum associated with the limit consists of the orientifold hyperplane given by $\zeta=0$, together with a singular brane supported on 
\[
D:(\eta^2+12\zeta^2 \vartheta=0)\subset Z.
\] 

\subsection{An SO(3) limit}
Let $X\to B$ be an SO(3) fibration as defined in \S\ref{VUC}, whose defining equation is given by
\[
y^2z=x^3+a_2x^2z+sxz^2.
\]
We then let 
\[
s=\sigma t, \quad a_2=h,
\]
where $t$ is a complex deformation parameter, varying over a disk $\mathscr{D}\subset \C$. As $t$ varies we then arrive at a family $\mathscr{X}\to \mathscr{D}$, whose central fiber is given by
\[
\mathscr{X}_0:(y^2z=x^3+hx^2z)\subset \mathbb{P}(\mathscr{E}).
\] 
We immediately see that the fibers of $\mathscr{X}_0$ over $h\neq 0$ are nodal curves, thus signaling weak coupling almost everywhere over $B$. Over $h=0$ the nodal curves then enhance to cuspidal curves, thus $h=0$ is identified with the orientifold hyperplane. The double cover $Z\to B$ corresponding to the orientifold is then constructed as exactly the same way as in Sen's limit, and the associated family of discriminants expanded with respect to $t$ is then given by
\[
\Delta(t)=h^2\sigma^2t^2+\cdots .
\]   
It then follows that the pullback to $Z$ of the flat limit of $\Delta(t)$ as $t\to 0$ is given by
\[
\overline{\Delta}_0: (\zeta^4\sigma^2=0)\subset Z,
\]
where we recall $\zeta$ is such that the equation for $Z$ is given by $\zeta^2=h$. We then see that the limiting brane spectrum in $Z$ corresponds to a smooth orientifold hyperplane given by $\zeta=0$, and a stack of 2 branes supported on the smooth divisor $\sigma=0$.

\subsection{An SO(5) limit}
Let $X\to B$ be an SO(5) fibration as defined in \S\ref{VUC}, whose defining equation is given by
\[
y^2z=x^3+a_2x^2z+s^2xz^2.
\]
We then let 
\[
s=\psi t, \quad a_2=h,
\]
where $t$ is a complex deformation parameter, varying over a disk $\mathscr{D}\subset \C$. As $t$ varies we then arrive at a family $\mathscr{X}\to \mathscr{D}$, whose central fiber is given by
\[
\mathscr{X}_0:(y^2z=x^3+hx^2z)\subset \mathbb{P}(\mathscr{E}).
\] 
We immediately see that the fibers of $\mathscr{X}_0$ over $h\neq 0$ are nodal curves, thus signaling weak coupling almost everywhere over $B$. Over $h=0$ the nodal curves then enhance to cuspidal curves, thus $h=0$ is identified with the orientifold hyperplane. The double cover $Z\to B$ given by $\zeta^2=h$ is then smooth, and the associated family of discriminants expanded with respect to $t$ is then given by
\[
\Delta(t)=h^2\psi^4t^4+\cdots .
\]   
It then follows that the pullback to $Z$ of the flat limit of $\Delta(t)$ as $t\to 0$ is given by
\[
\overline{\Delta}_0: (\zeta^4\psi^4=0)\subset Z.
\]
We then see that the limiting brane spectrum in $Z$ corresponds to a smooth orientifold hyperplane given by $\zeta=0$, and a stack of 4 branes supported on the smooth divisor $\psi=0$.

\subsection{An SO(6) limit}\label{SO6L}
Let $X\to B$ be an SO(6) fibration as defined in \S\ref{VUC}, whose defining equation is given by
\[
y^2z+a_1xyz=x^3+sx^2z+s^2xz^2.
\]
We then let 
\[
a_1=\zeta+\eta t, \quad s=\frac{1}{4}\zeta^2-\frac{1}{2}\zeta \eta t,
\]
which gives rise to a family $\mathscr{X}\to \mathscr{D}$, whose central fiber is given by
\[
\mathscr{X}_0:\left(y^2z+\zeta xyz=x^3+\frac{1}{4}\zeta^2 x^2z+\frac{1}{16}\zeta^4xz^2\right)\subset \mathbb{P}(\mathscr{E}).
\]
The Weierstrass coefficients of the central fiber are then given by
\[
F=-\frac{1}{48}\zeta^4, \quad G=-\frac{1}{864}\zeta^6.
\]
We immediately see that the fibers of $\mathscr{X}_0$ over $\zeta\neq 0$ are nodal curves, thus signaling weak coupling almost everywhere over $B$. Over $\zeta=0$ the nodal curves then enhance to cuspidal curves. We then define a double cover $Z\to B$, whose total space is given by $\zeta^2=h$, where $h$ is a general section of $\mathscr{L}^2$. The total space of the double cover $Z\to B$ is then smooth, and the associated family of discriminants expanded with respect to $t$ is then given by
\[
\Delta(t)=\zeta^{11}\eta t+\cdots .
\]   
It then follows that the pullback to $Z$ of the flat limit of $\Delta(t)$ as $t\to 0$ is given by
\[
\overline{\Delta}_0: (\zeta^{11}\eta=0)\subset Z.
\]
Viewing the equation of $\overline{\Delta}_0$ as $\zeta^4\zeta^{7} \eta=0$, we then see that the limiting brane spectrum in $Z$ corresponds to a smooth orientifold hyperplane given by $\zeta=0$, a stack of 7 branes supported on the orientifold hyperplane plane, and a single brane supported on the smooth divisor $\eta=0$. If we specialize the situation by letting $\eta=\zeta$, then the brane spectrum corresponds to a stack of 4 brane-image-brane pairs supported on the orientifold hyperplane.

\subsection{An $F_4$ limit}
Let $X\to B$ be an $F_4$ fibration as defined in \S\ref{VUC}, whose defining equation is given by
\[
y^2z=x^3+c_1s^3xz^2+c_2s^4z^3.
\]
We then let 
\[
s=\zeta, \quad c_1=-3\zeta, \quad c_2=2\zeta^2+\eta^2 t
\]
which gives rise to a family $\mathscr{X}\to \mathscr{D}$, whose central fiber is given by
\[
\mathscr{X}_0:\left(y^2z=x^3-3\zeta^4 xz^2+2\zeta^6z^3\right)\subset \mathbb{P}(\mathscr{E}).
\]
The Weierstrass coefficients of the central fiber are then given by
\[
F=-3\zeta^4, \quad G=2\zeta^6.
\]
We immediately see that the fibers of $\mathscr{X}_0$ over $\zeta\neq 0$ are nodal curves, thus signaling weak coupling almost everywhere over $B$. Over $\zeta=0$ the nodal curves then enhance to cuspidal curves. We then define a double cover $Z\to B$, whose total space is given by $\zeta^2=h$, where $h$ is a general section of $\mathscr{L}^2$. The total space of the double cover $Z\to B$ is then smooth, and the associated family of discriminants expanded with respect to $t$ is then given by
\[
\Delta(t)=4\zeta^{10}\eta^2t+\cdots .
\]   
It then follows that the brane spectrum is exactly the same as in the SO(6) limit constructed in \S\ref{SO6L}, and as such, each of the SO(6) and $F_4$ fibrations may be viewed as distinct F-theory lifts of this orientifold limit.

\subsection{An SU(4) limit}\label{SU4L}
Let $X\to B$ be an SU(4) fibration as defined in \S\ref{VUC}, whose defining equation is given by
\[
y^2z+a_1xyz=x^3+c_1sx^2z+c_2s^2xz^2+d_2s^4z^3.
\]
We then let 
\[
a_1=\zeta, \quad s=\gamma, \quad d_2=\eta t^3,\quad c_1=\alpha t, \quad c_2=\beta t,
\]
which gives rise to a family $\mathscr{X}\to \mathscr{D}$, whose central fiber is given by
\[
\mathscr{X}_0:\left(y^2z+\zeta xyz=x^3\right)\subset \mathbb{P}(\mathscr{E}).
\]
The Weierstrass coefficients of the central fiber are then given by
\[
F=-\frac{1}{48}\zeta^4, \quad G=\frac{1}{864}\zeta^6.
\]
We immediately see that the fibers of $\mathscr{X}_0$ over $\zeta\neq 0$ are nodal curves, thus signaling weak coupling almost everywhere over $B$. Over $\zeta=0$ the nodal curves then enhance to cuspidal curves. We then define a double cover $Z\to B$, whose total space is given by $\zeta^2=h$, where $h$ is a general section of $\mathscr{L}^2$. The total space of the double cover $Z\to B$ is then smooth, and the associated family of discriminants expanded with respect to $t$ is then given by
\[
\Delta(t)=-\zeta^4\gamma^4\beta^2t^4+\cdots .
\]   
It then follows that the pullback to $Z$ of the flat limit of $\Delta(t)$ as $t\to 0$ is given by
\[
\overline{\Delta}_0: (\zeta^{4}\gamma^4\beta^2=0)\subset Z.
\]
We then see that the brane spectrum in $Z$ corresponds to a smooth orientifold hyperplane given by $\zeta=0$, a stack of 4 branes supported on the smooth divisor $\gamma=0$, and a stack of 2 branes supported on the smooth divisor $\beta=0$. If we specialize the situation by letting $\gamma=\zeta$, then the stack of 4 branes supported on $\gamma=0$ then becomes a stack of 2 brane-image-brane pairs supported on the orientifold hyperplane.

\subsection{A Spin(7) limit}
Let $X\to B$ be a Spin(7) fibration as defined in \S\ref{VUC}, whose defining equation is given by
\[
y^2z=x^3+c_1sx^2z+c_2s^2xz^2+d_2s^4z^3.
\]
We then let 
\[
s=\zeta, \quad c_1=\zeta \quad d_2=\eta t^3,\quad c_2=\beta t,
\]
which gives rise to a family $\mathscr{X}\to \mathscr{D}$, whose central fiber is given by
\[
\mathscr{X}_0:\left(y^2z+\zeta xyz=x^3\right)\subset \mathbb{P}(\mathscr{E}).
\]
The Weierstrass coefficients of the central fiber are then given by
\[
F=-\frac{1}{3}\zeta^4, \quad G=\frac{2}{27}\zeta^6.
\]
We immediately see that the fibers of $\mathscr{X}_0$ over $\zeta\neq 0$ are nodal curves, thus signaling weak coupling almost everywhere over $B$. Over $\zeta=0$ the nodal curves then enhance to cuspidal curves. We then define a double cover $Z\to B$, whose total space is given by $\zeta^2=h$, where $h$ is a general section of $\mathscr{L}^2$. The total space of the double cover $Z\to B$ is then smooth, and the associated family of discriminants expanded with respect to $t$ is then given by
\[
\Delta(t)=-\zeta^8\beta^2t^4+\cdots .
\]   
It then follows that the pullback to $Z$ of the flat limit of $\Delta(t)$ as $t\to 0$ is given by
\[
\overline{\Delta}_0: (\zeta^8\beta^2=0)\subset Z.
\]
We then see that the brane spectrum in $Z$ is precisely the same as the SU(4) limit constructed in \S\ref{SU4L} specialized at $\gamma=\zeta$, and as such, each of the SU(4) and Spin(7) fibrations may be viewed as distinct $F$-theory lifts of this limit.

\section{Universal tadpole relations}
As D3 charge is preserved under S-duality, the D3 tadpole of a consistent, global orientifold limit of $F$-theory should coincide with that of its $F$-theory lift. For a smooth $F$-theory compactification $X\to B$, the D3 tadpole is given by
\[
N_{D3}=\frac{1}{24}\chi(X),
\]
while on the type-IIB side the D3 tadpole is given by
\[
N_{D3}=\frac{1}{2}\left(4\sum_i \frac{\chi(O_i)}{24}+\sum_j\frac{\chi(D_j)}{24}\right),
\]
where the $O_i$ are the supports of the orientifold hyperplanes and the $D_j$ are the supports of the codimension 1 branes in the orientifold double cover $Z\to B$. Equating the two tadpoles then yields the consistency condition
\begin{equation}\label{TR1}
2\chi(X)=4\sum_i \chi(O_i)+\sum_j \chi(D_j).
\end{equation}
In each of the limits constructed in \S\ref{OL}, the brane spectrum consists of a single orientifold hyperplane $O$ together with branes supported on smooth divisors in $Z$, and in each case, what we find is that the tadpole relation \eqref{TR1} holds if and only if the Euler characteristic $\chi(X)$ appearing on the LHS of the relation is replaced with the \emph{stringy} Euler characteristic $\chi_{\text{str}}(X)$. In particular, for each limit constructed in \S\ref{OL}, we find
\begin{equation}\label{STR}
2\chi_{\text{str}}(X)=4\chi(O)+\sum_j \chi(D_j).
\end{equation}
We take this as strong evidence that for singular $F$-theory compactifications, the D3 tadpole should be given by
\[
N_{D3}=\frac{1}{24}\chi_{\text{str}}(X).
\]
Moreover, in each of the limits constructed in \S\ref{OL}, we find that equation \eqref{STR} is the dimension zero component of the homological identity
\begin{equation}\label{SUTR}
2\varphi_*c_{\text{str}}(X)=\rho_*(4c(O)+\sum_j c(D_j)),
\end{equation}
where $\varphi:X\to B$ is the $F$-theory compactification and $\rho:Z\to B$ is the associated orientifold compactification. 

Stringy Chern classes are defined for varieties $X$ with at most Gorenstein canonical singularities \cite{SCCA}\cite{SCCdF}, and in the case that $X$ admits a crepant resolution $\tau:\widetilde{X}\to X$, so that $\tau^*K_X=K_{\widetilde{X}}$, the definition of stringy Chern class yields $\tau_*c(\widetilde{X})=c_{\text{str}}(X)$. Stringy Chern classes reside in the Chow group of algebraic cycles modulo rational equivalence, and the stringy version of the Gauss-Bonnet theorem is then given by the formula
\[
\chi_{\text{str}}(X)=\int_X c_{\text{str}}(X),
\]
where the integral sign in notation for taking the degree of the zero-dimensional component of a Chow class. 

For each singular $F$-theory compactification $\varphi:X\to B$ we consider, crepant resolutions $\tau:\widetilde{X}\to X$ were constructed in \cite{EJK}, which were then used to compute explicit formulas for $\varphi_*(\tau_*c(\widetilde{X}))$ (which may also be recovered by taking the limit as $y\to -1$ of the stringy Hirzebruch class formulas derived in \cite{SHCFvH}). And since $\tau_*c(\widetilde{X})=c_{\text{str}}(X)$, we have $\varphi_*(\tau_*c(\widetilde{X}))=\varphi_*c_{\text{str}}(X)$, which will enable us to compute the LHS of the universal tadpole relation \eqref{SUTR} associated with each limit. As the degree of a zero-dimensional Chow class is invariant under proper pushforward \cite{IT}, we also have
\begin{equation}\label{SGB}
\chi_{\text{str}}(X)=\int_B \varphi_*c_{\text{str}}(X),
\end{equation}
thus the dimension-zero component of \eqref{SUTR} encodes the numerical identity \eqref{STR} corresponding to the matching of the D3 tadpoles associated with an orientifold limit of $F$-theory. 

In what follows, we verify the universal tadpole relation \eqref{SUTR} associated with each limit constructed in \S\ref{OL}. In each case, we denote the first Chern class of $\mathscr{L}\to B$ by $L$, and we will use the fact that if $D_a$ is a smooth divisor of class $\rho^*aL$ in the total space of the orientifold $\rho:Z\to B$, then
\begin{equation}\label{CSD}
\rho_*c(D_a)=\frac{2aL}{1+aL}\cdot \frac{1+L}{1+2L}c(B).
\end{equation} 
For explicit verifications of formula \eqref{CSD} for various $a$, one may consult \S4 of \cite{AE2}.

\subsection{The SO(3) universal tadpole relation}
Let $\varphi:X\to B$ be an SO(3) fibration. As computed in \cite{EJK}, we have 
\[
\varphi_*c_{\text{str}}(X)=\frac{12L}{1+4L}c(B).
\]
The brane spectrum in the total space of $\rho:Z\to B$ associated with the orientifold limit of SO(3) fibrations constructed in \S\ref{OL} consists of a single orientifold hyperplane supported on a smooth divisor $O$ given by $\zeta=0$ and a stack of two branes supported on a smooth divisor $D$ given by $\sigma=0$. Since $[O]=\rho^*L$ and $[D]=\rho^*4L$, formula \eqref{CSD} yields
\[
\rho_*c(O)=\frac{2L}{1+2L}c(B) \quad \text{and} \quad \rho_*c(D)=\frac{8L}{1+4L}\cdot \frac{1+L}{1+2L}c(B).
\]
We then have
\begin{eqnarray*}
\rho_*(4c(O)+2c(D))&=&\left(4\frac{2L}{1+2L}+2\frac{8L}{1+4L}\cdot \frac{1+L}{1+2L}\right)c(B) \\
&=&\left(\frac{8L(3+6L)}{(1+2L)(1+4L)}\right)c(B) \\
&=&\frac{24L}{1+4L}c(B) \\
&=&2\varphi_*c_{\text{str}}(X), \\
\end{eqnarray*}
thus $2\varphi_*c_{\text{str}}(X)=\rho_*(4c(O)+2c(D))$, as desired.

\subsection{The SO(5) universal tadpole relation}
Let $\varphi:X\to B$ be an SO(5) fibration. As computed in \cite{EJK}, we have 
\[
\varphi_*c_{\text{str}}(X)=\frac{4L(3+4L)}{(1+2L)^2}c(B).
\]
The brane spectrum in the total space of $\rho:Z\to B$ associated with the orientifold limit of SO(5) fibrations constructed in \S\ref{OL} consists of a single orientifold hyperplane supported on a smooth divisor $O$ given by $\zeta=0$, and a stack of 4 branes supported on a smooth divisor $D$ given by $\psi=0$. Since $[O]=\rho^*L$ and $[D]=\rho^*2L$, we have
\[
\rho_*c(O)=\frac{2L}{1+2L}c(B) \quad \text{and} \quad \rho_*c(D)=\frac{4L}{1+2L}\cdot \frac{1+L}{1+2L}c(B).
\]
We then have
\begin{eqnarray*}
\rho_*(4c(O)+4c(D))&=&\left(4\frac{2L}{1+2L}+4\frac{4L}{1+2L}\cdot \frac{1+L}{1+2L}\right)c(B) \\
&=&\frac{8L(3+4L)}{(1+2L)^2}c(B) \\
&=&2\varphi_*c_{\text{str}}(X), \\
\end{eqnarray*}
thus $2\varphi_*c_{\text{str}}(X)=\rho_*(4c(O)+4c(D))$, as desired.

\subsection{The SO(6) and $F_4$ universal tadpole relations}
Let $\varphi:X\to B$ be an SO(6) or an $F_4$ fibration. As computed in \cite{EJK}, in both cases we have 
\[
\varphi_*c_{\text{str}}(X)=\frac{12L}{1+2L}c(B).
\]
The brane spectrum in the total space of $\rho:Z\to B$ associated with the orientifold limit of $F_4$ fibrations constructed in \S\ref{OL} consists of a single orientifold hyperplane supported on a smooth divisor $O$ given by $\zeta=0$, a stack of 6 branes supported on $O$ (which may be viewed as brane-image-brane pairs), and a stack of two branes supported on a smooth divisor $D$ given by $\eta=0$. Since $[O]=[D]=\rho^*L$, we have
\[
\rho_*c(O)=\rho_*c(D)\frac{2L}{1+2L}c(B).
\]
We then have
\begin{eqnarray*}
\rho_*(4c(O)+6c(O)+2c(D))&=&\left(4\frac{2L}{1+2L}+6\frac{2L}{1+2L}+2\frac{2L}{1+2L}\right)c(B) \\
&=&\frac{24L}{1+2L}c(B) \\
&=&2\varphi_*c_{\text{str}}(X), \\
\end{eqnarray*}
thus $2\varphi_*c_{\text{str}}(X)=\rho_*(10c(O)+2c(D))$, as desired. In the SO(6) case, instead of having 6 branes supported on $O$ and 2 branes supported on $D$, we have 7 branes supported on $O$ and 1 brane supported on $D$. And since $[O]=[D]$, the tadpole relation holds just as in the $F_4$ case. We note that this universal tadpole relation associated with the SO(6) and $F_4$ limits provide geometric realizations of a universal tadpole relation that was shown to exist in \cite{AE2} (i.e., the $(1,...,1)$ case of Theorem~4.9). To the best of our knowledge these are the first such geometric realizations of this universal tadpole relation. 

\begin{remark}
In the SO(6) and $F_4$ cases there is a strongly coupled matter sector localized along the codimension 2 locus $\eta=\zeta=0$ due to the order of vanishing of the Weierstrass coefficients $F$ and $G$ upon this locus. As such, one would expect a contribution to the D3 tadpole coming from $\eta=\zeta=0$, but we find the the usual tadpole relation as in the Sen limit still holds nonetheless. Perhaps this is due to the fact that the \emph{stringy} Euler characteristic -- which we are using for the D3 tadpole on the $F$-theory side -- `sees' the extra contribution coming from the codimension 2 locus $\eta=\zeta=0$. In any case, a more precise explanation for the usual tadpole relation holding in the SO(6) and $F_4$ cases in light of the presence of the strongly coupled localized matter sector would certainly be desirable.
\end{remark}

\subsection{The SU(4) and Spin(7) universal tadpole relations}
Let $\varphi:X\to B$ be an SU(4) or a Spin(7) fibration. As computed in \cite{EJK}, in both cases we have 
\[
\varphi_*c_{\text{str}}(X)=\frac{4L(3+5L)}{(1+2L)^2}c(B).
\]
The brane spectrum in the total space of $\rho:Z\to B$ associated with the SU(4) limit constructed in \S\ref{OL} consists of a single orientifold hyperplane supported on a smooth divisor $O$ given by $\zeta=0$, a stack of 4 branes supported on a smooth divisor $D_1$ given by $\gamma=0$, and a stack of 2 branes supported on a smooth divisor $D_2$ given by $\beta=0$. The spectrum in the Spin(7) limit then corresponds to setting $\gamma=\zeta$, which makes no difference in the tadpole relation since $[D_1]=[O]$. And since $[D_2]=\rho^*2L$, we have
\[
\rho_*c(O)=\rho_*c(D_1)\frac{2L}{1+2L}c(B) \quad \text{and} \quad \rho_*c(D_2)=\frac{4L}{1+2L}\cdot \frac{1+L}{1+2L}c(B).
\]
We then have
\begin{eqnarray*}
\rho_*(4c(O)+4c(D_1)+2c(D_2))&=&\left(4\frac{2L}{1+2L}+4\frac{2L}{1+2L} +2\frac{4L}{1+2L}\cdot \frac{1+L}{1+2L}\right)c(B) \\
&=&\frac{8L(3+5L)}{(1+2L)^2}c(B) \\
&=&2\varphi_*c_{\text{str}}(X), \\
\end{eqnarray*}
thus $2\varphi_*c_{\text{str}}(X)=\rho_*(4c(O)+4c(D_1)+2c(D_2))$, as desired.

\section{Tadpole cancellation with smooth branes}\label{SSB}
If an orientifold limit of an $F$-theory compactification $\varphi:X\to B$ consists of a configuration of $k$ smooth branes in the total space of the orientifold $\rho:Z\to B$, equation \eqref{CSD} implies that the expected universal tadpole relation associated with the limit takes the form 
\begin{equation}\label{utr17}
2\varphi_*c_{\text{str}}(X)=\left(\sum_{i=1}^k\frac{2a_i L}{1+a_iL}\right)\frac{1+L}{1+2L}c(B).
\end{equation}
Further assuming that $X$ is given by the vanishing of a Tate form implies that the discriminant of $\varphi:X\to B$ has divisor class $12L$, which constrains the $a_i$s in \eqref{utr17} to be such that $\sum a_i=12$. This constraint along with the fact that $2\varphi_*c_{\text{str}}(X)/c(B)$ is a rational function in $L$ implies that if a universal tadpole relation of the form \eqref{utr17} exists, then it is necessarily unique. Moreover, one may reverse engineer orientifold limits with smooth branes by imposing equation \eqref{utr17} as a constraint for the limit. From such a perspective one may also use equation \eqref{utr17} to show an orientifold limit satisfying a universal tadpole relation doesn't exist. For example, an SU(5) fibration is given by the equation
\[
\text{SU(5)}: y^2z + a_1xyz + b_1s^2yz^2=x^3 + c_1sx^2z + d_1s^{3}xz^2 + e_1s^{5}z^3,
\]
and the pushforward of its stringy Chern class is given by 
\[
\varphi_*c_{\text{str}}(X)=\frac{7L^3+14L^2+12L}{(1+L)^3}c(B).
\]
Equation \eqref{utr17} then implies that an SU(5) limit with solely smooth branes admits a universal tadpole relation if and only if there exists $a_i\in \mathbb{N}$ (with $\sum a_i=12$) such that
\begin{equation}\label{utrsu5}
\frac{7L^3+14L^2+12L}{(1+L)^3} = \frac{1+L}{1+2L} \sum_i \frac{a_iL}{1+a_iL}.
\end{equation}
But equation \eqref{utrsu5} may be re-written as
\[
(7L^3+14L^2+12L)(1+2L)= \sum_i \left(\frac{a_iL}{1+a_iL}(1+L)^{4}\right),
\]
and a substitution of $L= -1$, shows that no such $a_i\in \mathbb{N}$ exist. As such, there can be no orientifold limit of an SU(5) fibration consisting of a purely smooth brane spectrum which also admits a universal tadpole realtion. It then follows that if an orientifold limit of an SU(5) admits a universal tadpole relation, the limit must necessarily contain singular branes, as in Sen's limit for smooth Weierstrass fibrations.

\end{document}